\documentclass[preprint,showpacs,preprintnumbers,amsmath,amssymb,superscriptaddress]{revtex4}

\usepackage{graphicx}
\usepackage{dcolumn}
\usepackage{bm}

\begin{document}

\preprint{APS/123-QED}

\title{Study of ortho-to-paraexciton conversion in Cu$_2$O by excitonic Lyman spectroscopy}

\author{M. Kubouchi}
\affiliation{%
Department of Applied Physics, the University of Tokyo, and 
Solution Oriented Research for Science and Technology (SORST), JST, 7-3-1 Hongo, Bunkyo-ku, Tokyo 113-8656, Japan
}

\author{K. Yoshioka}
\affiliation{%
Department of Applied Physics, the University of Tokyo, and 
Solution Oriented Research for Science and Technology (SORST), JST, 7-3-1 Hongo, Bunkyo-ku, Tokyo 113-8656, Japan
}

\author{R. Shimano}
\altaffiliation{present address: Department of Physics, The University of Tokyo, Tokyo 113-8656, Japan}
\affiliation{%
Department of Applied Physics, the University of Tokyo, and 
Solution Oriented Research for Science and Technology (SORST), JST, 7-3-1 Hongo, Bunkyo-ku, Tokyo 113-8656, Japan
}

\author{A. Mysyrowicz}
\affiliation{%
Laboratoire d'Optique Appliquee, ENSTA, Ecole Polytechnique, Palaiseau, France
}%

\author{M. Kuwata-Gonokami}
\email{gonokami@ap.t.u-tokyo.ac.jp}
\affiliation{%
Department of Applied Physics, the University of Tokyo, and 
Solution Oriented Research for Science and Technology (SORST), JST, 7-3-1 Hongo, Bunkyo-ku, Tokyo 113-8656, Japan
}

\date{\today}

\begin{abstract}
Using time-resolved $1s$-$2p$ excitonic Lyman spectroscopy,
 we study the orthoexciton-to-paraexcitons transfer,
 following the creation of a high density population of ultracold $1s$ orthoexcitons 
by resonant two-photon excitation with femtosecond pulses. 
 An observed fast exciton-density dependent conversion rate is attributed to spin exchange between
 pairs of orthoexcitons.
  Implication of these results on the feasibility of BEC of paraexcitons in Cu$_2$O is discussed. 
\end{abstract}

\pacs{71.35.Lk, 71.35.Cc, 78.47.+p}
\maketitle
The observation of Bose-Einstein condensation (BEC) of neutral atoms, more than 70 years after its prediction, 
constitutes a major advance in physics of the last decade \cite{rb,ketterle}. 
Ensembles of ultracold atoms with high controllability of density, 
temperature, and interaction strength reveals new aspects of many body quantum phenomena. 
In particular, by applying a magnetic field one can control the sign and 
magnitude of the scattering length between atoms. 
This allows probing the low temperature transition between collective pairing of fermionic atoms 
with attractive interaction and BEC of molecularlike bosonic entities \cite{jin}.  
Excitons, composite particles in semiconductors made of fermions, may provide another system 
particularly well suited for the study of this transition in a slightly different context. 
With increasing densities, the fluid should evolve continuously from 
a dilute Bose-Einstein condensate of excitons into a dielectric superfluid consisting 
of a BCS-like degenerate two-component Fermi system with Coulomb attraction\cite{keldysh,conte}. 
Several recent experimental results have renewed interest in the search of 
BEC in photo excited semiconductors\cite{butov}. 

It has been long recognized that Cu$_2$O provides a material with unique advantages for 
the observation of BEC of excitons \cite{mysyrowicz}. Because of the positive parity of the valence 
and conduction band minima at the center of the Brillouin zone, their radiative recombination is 
forbidden in the dipole approximation, conferring a long radiative lifetime to the $n=1$ exciton. 
The $n=1$ exciton level is split by exchange interaction in a triply degenerate orthoexciton state 
(symmetry $\Gamma _5^+$; 2.033 eV at 2 K), 
and a lower lying singly degenerate paraexciton ($\Gamma _2^+$; 2.021 eV at 2 K) 
which is optically inactive to all orders. Several experiments have shown intriguing results in luminescence 
and transport suggesting the occurrence of a degenerate excitonic quantum fluid at high densities and low temperatures 
in this material \cite{fortin,snoke}. However, a controversy has been raised on the actual density of excitonic particles 
created by optical pumping. Based on luminescence absolute quantum efficiency measurement, 
it has been claimed that a Auger effect with giant cross section destroys excitons 
when the density exceeds 10$^{15}$ cm$^{-3}$, preventing BEC \cite{ohara}. On the other hand, 
other authors have shown that another process, orthoexciton-paraexciton conversion by spin exchange 
is much more probable at low temperature \cite{prb61_16619}. Since this last effect does not destroy excitons, 
but merely increases the orthoexciton-paraexciton conversion rate, it does not prevent BEC. 
To resolve this controversy, it is therefore of prime importance to explore the ortho-para conversion rate 
as a function of density. More generally it is important to study the dynamics of paraexcitons 
in order to optimize the pumping conditions to achieve BEC. 
One of the major difficulties in this respect was the lack up to now of a sensitive spectroscopic method 
to probe optically inactive paraexcitons. 

In this Letter, we use a new spectroscopic approach to study the dynamics 
of orthoexciton-paraexciton conversion in Cu$_2$O. The technique consists of probing the transition from 
the populated $1s$ to the $2p$ state, with a midinfrared (MIR) light beam. 
MIR excitonic spectroscopy is the counterpart of Lyman spectroscopy in atomic hydrogen \cite{haken,jcs45_949,johnsen}. 
It is especially well suited for the study of paraexcitons in Cu$_2$O because the $1s$-$2p$ transition 
is allowed even if the $1s$ paraexciton is optically inactive. With the use of a short pump pulse 
to excite orthoexciton and a weak MIR short probe light pulse, one can follow the gradual buildup 
of the paraexciton population from ortho-para conversion and its subsequent decay. 
Since the dipole matrix element for $1s$-$2p$ transition is known, 
one can extract the density of paraexcitons from the strength of the Lyman absorption. 
This has to be contrasted with luminescence data, where exciton density estimates 
rely on measurements of absolute radiative quantum efficiency, a notoriously difficult task. 
In addition, the lineshape of Lyman absorption yields precise information on the energy distribution of 
$1s$ excitons. In particular, Johnsen and Kavoulakis pointed out that is should show a characteristic 
abrupt change when excitonic BEC occurs \cite{johnsen}.

To follow the density and temperature evolution of the orthoexcitons and paraexcitons starting with a well controlled situation, 
we first generate orthoexcitons 
by resonant two-photon absorption (TPA), using an ultrashort laser pulse. Drawing an analogy 
with the two-photon excitation of biexcitons in CuCl with ultrashort laser pulses \cite{cucl}, 
we note that the created orthoexcitons have a very low initial temperature despite the large laser bandwidth, 
because of the small group velocity dispersion at the TPA laser frequency ($\hbar \omega =1.0164$ eV). 
In fact, the initial momentum spread of the generated orthoexcitons is even much smaller 
than in the case of biexcitons in CuCl. A conservative estimate yields an initial orthoexciton temperature 
much less than $10^{-3}$ K.

\begin{figure}
\includegraphics{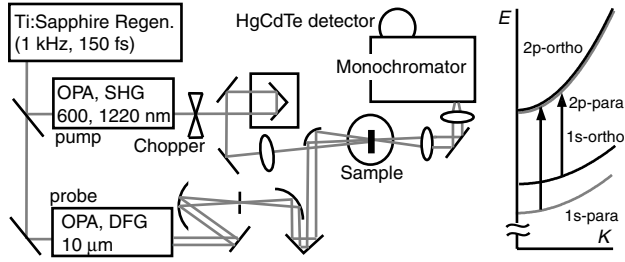}
\caption{\label{figure1} Experimental set up for time-resolved pump-probe spectroscopy of 
excitonic Lyman transitions. The 150 fs duration pump pulse is obtained 
from a Ti:sapphire laser and optical parametric amplifier. 
The pump pulse wavelength can be tuned around 1220 or 600 nm. The midinfrared probe pulse, 
of same duration is obtained by parametric down conversion. It is tunable around 10 $\mu$m. 
On the right hand side, the energy diagram of the relevant levels is shown.}
\end{figure}

\begin{figure}
\includegraphics{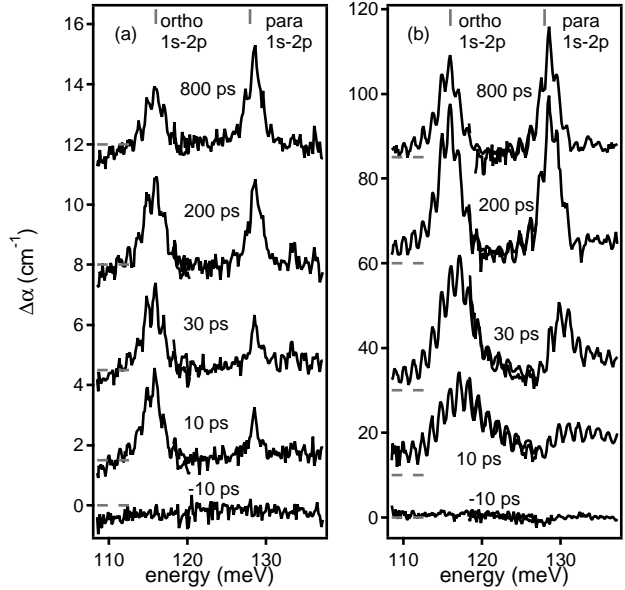}
\caption{\label{figure2}
(a) Lyman absorption recorded in a 170 $\mu$m thick single crystal of Cu$_2$O held at 4.2 K 
at different delays from the pump beam. The pump beam tuned at 1220 nm 
(two-photon resonant excitation of the orthoexciton) is incident along a [100] crystal axis.
(b) Same as in (a), except for the pump wavelength, now tuned at 600 nm 
(non-resonant one photon excitation of orthoexcitons).}
\end{figure}

The experimental setup and a relevant energy diagram are shown in Fig. \ref{figure1}. 
A 150 fs laser pulse tunable around 1220 nm provides the pump source. 
A tunable light pulse in the MIR around 10 $\mu$m provides the weak probe source. 
The 170 $\mu$m thick single crystal platelets were cooled by contact with a liquid Helium bath. 
Unless otherwise specified, the pump laser was propagating along a [100] crystal axis.

Representative Lyman absorption spectra are shown in Fig. \ref{figure2} at various 
pump-probe delays ranging between $-10$ and 800 ps. Figure \ref{figure2}(a) is obtained 
with two-photon resonant excitation using a pump pulse energy of 1.0 $\mu$J 
focused on a spot diameter of 400 $\mu$m (0.81 mJ/cm$^2$), 
corresponding to an intensity of 5.4 GW/cm$^2$ and an esimated initial orthoexciton density of about 10$^{16}$ cm$^{-3}$ 
assuming a two-photon absorption coefficient of $\beta = 0.001$ cm/MW \cite{jolk}.
Figure \ref{figure2}(b) is obtained by tuning the pump pulse to 600 nm, 
inside the phonon-assisted absorption edge of the $n=1$ orthoexciton. 
Because the exciton hyperfine splitting is large (12 meV) for the $n=1$ level and negligibly 
small (much smaller than the experimental limit of the spectral resolution 0.3 meV)
 for the $n=2$ level, one can record simultaneously the Lyman transition for 
orthoexcitons and paraexcitons. 
Under both pumping conditions, two lines appear around 116 meV and 129 meV, exactly
where the $1s$-$2p$ Lyman transitions of the ortho- and paraexcitons are expected \cite{kubouchi}.

\begin{figure}
\includegraphics{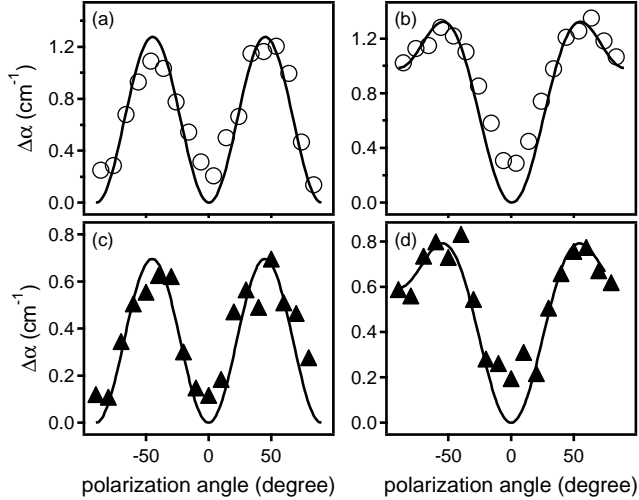}
\caption{\label{figure3} Lyman absorption of orthoexciton (open circles) measured 
as a function of angle between the laser polarization and the crystal axis [001], 
for two crystal orientations, $\boldsymbol{k}\parallel$ [100] (a) and $\boldsymbol{k}\parallel$ [110] (b). 
The Lyman absorption of paraexciton [closed triangles: (c),(d)] exhibits the same dependence.}
\end{figure}

We have carefully verified that the appearance of a signal at the position of the paraexciton line at 129 meV 
in  Fig. \ref{figure2}(a) is not simply due to the direct creation of paraexcitons by the pump pulse, 
using the following measurements.
First, it was verified that both lines at 116 meV and 129 meV disappear 
if the pump frequency is tuned off orthoexciton resonance.
Secondly, we have measured their polarization dependence as a function of the pump polarization vector, 
for two different direction of propagation with respect to the crystal axes (see Fig. \ref{figure3}).
As mentioned before, the two-photon absorption to the paraexciton is forbidden 
so that no polarization dependence can be assigned to such a transition. 
One expect, for the orthoexciton state $\Gamma_5^+$, 
a dependence of the two-photon absorption with the polarization angle of the form \cite{inoue,bader}:
\begin{eqnarray}
\Delta \alpha \propto \sin ^2 2\theta,
\end{eqnarray} if $\boldsymbol{k}\parallel$ [100], and
\begin{eqnarray}
\Delta \alpha \propto \sin ^2 2\theta + \sin ^4 \theta,
\end{eqnarray} for $\boldsymbol{k}\parallel$ [110].
Where $\Delta \alpha$ is the induced absorption, $\theta$ is the angle between the polarization vector 
and the crystal axis [001] and the $\boldsymbol{k}$ vector points along the laser beam direction. 
The observed behaviour for both lines at 116 meV and 129 meV are the same, as shown in Fig. \ref{figure3}, 
and it exhibits the polarization dependence expected for a two-photon transition to the $\Gamma_5^+$ orthoexciton state. 
Both experiments therefore indicate that orthoexcitons are first created by the pump source,
and subsequently decay into paraexcitons.

The temporal evolution of the shapes of the two Lyman absorption lines shown in Fig. \ref{figure2} 
reflects the dynamics of the distribution functions of orthoexcitons and paraexcitons. 
At long delay, $\Delta t >$ 200 ps, the lines acquire a width corresponding to the lattice temperature 
both under one-photon and two-photon pumping. 
At early times, however, there is a significant difference, depending on the type of 
excitation (one or two photon). In the one-photon excitation, the lines are broader and shifted to 
higher energies [see Fig. \ref{figure2}(b)]. This signals a higher excitonic effective temperature, 
due to the excess energy 
delivered to the exciton gas in the pumping process. With resonant two-photon excitation [see Fig. \ref{figure2}(a)],
the phase space compression scheme confers an initial effective temperature 
to the orthoexciton gas well below that of the lattice, as already mentioned \cite{cucl}. 

\begin{figure}
\includegraphics{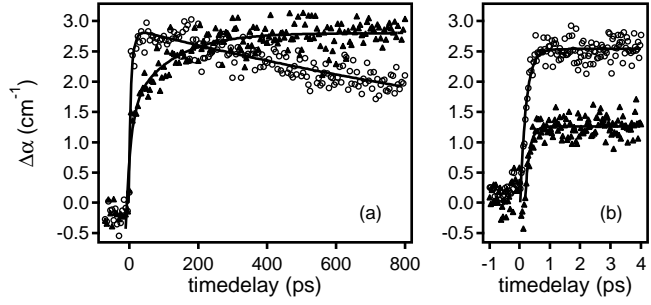}
\caption{\label{figure4} (a) Kinetics of the Lyman absorption of ortho- (open circles) and paraexcitons (closed triangles)
following two-photon resonant excitation of the orthoexciton line. 
(b) The expanded trace at early time shows the fast rise of the paraexciton population. 
Crystal temperature is 4.2 K.
The excitation density is 0.81 mJ/cm$^2$}
\end{figure}

\begin{figure}
\includegraphics{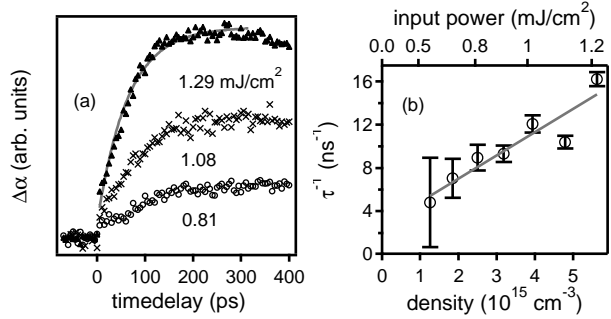}
\caption{\label{figure5} (a) Variation of the increase of the Lyman absorption of 
paraexcitons for various pump intensities. The increase of absorption is fitted 
to an exponential, as shown. 
(b) The exponential values obtained at different pump laser intensities are shown. 
The line obeys the relation: $\tau^{-1}=a + C_{\text{exp}} n_{\text{ex}}$, 
	with $\ a=2.7 \pm 1.4 \ \mathrm{ns}^{-1}$ and 
	$C_{\text{exp}}= (2.8 \pm 0.4)\times 10^{-15}\ \mathrm{cm}^{3}$/ns.}
\end{figure}

The time evolution of the ortho- and paraexcitons densities is shown in more details 
within a small [Fig. \ref{figure4}(b)] and large [Fig. \ref{figure4}(a)] time interval under two-photon resonant 
excitation of $1s$ orthoexcitons with a pump pulse energy of 0.81 mJ/cm$^2$. One can observe a very fast rise 
of the line at 129 meV, in less than 1 ps.
A different kinetics is observed on a longer time scale. The line population keeps increasing, 
but at a slower rate while the orthoexciton line at 116 meV decays with a time constant of the order of 1 ns. 
Figure \ref{figure5} shows the variation of the paraexciton Lyman absorption at 129 meV on a long time scale 
at higher pump intensities. Its rise time in the interval 2-100 ps becomes 
faster when the pump intensity is increased [Fig. \ref{figure5}(a)]. It can be approximately fitted 
by an exponential, with a laser intensity dependent time constant, as shown in Fig. \ref{figure5}(b).

To discuss the kinetics, we distinguish three characteristic times: picoseconds, 100 ps, and nanoseconds. 
We start with the slowest process which have been recently discussed comprehensively \cite{jang}. 
From the temperature dependence of the exciton luminescence kinetics, 
Jang \textit{et al.} have shown that ortho-para conversion occurs via the participation of a transverse
acoustic phonon.
This mechanism is relatively slow, with a characteristic conversion time of the order of nanoseconds. 
It is independent of particle density but increases with temperature.  
The slow decay rate of orthoexcitons with a nanoseconds time constant seen in Fig. \ref{figure4}, 
accompanied by a buildup of paraexcitons at a similar rate is consistent with this process.

We next consider 
behaviour in the range of 100 ps, where we observe the excitation density dependent 
increase of the area of the paraexciton Lyman line as shown in Fig. \ref{figure5}. 
 We can exclude Auger-type process with loss of particles since 
the area of orthoexciton and paraexciton signals are almost conserved as we find in Fig. \ref{figure4}.  
Kavoulakis and Mysyrowicz have proposed a fast orthoexciton-paraexciton conversion effective 
at high exciton densities and low temperature \cite{prb61_16619}.
It corresponds to an electron spin exchange between 
two orthoexcitons in a relative singlet spin configuration, resulting in their conversion in two paraexcitons. 
Such a mechanism scales like 
\begin{eqnarray}
\frac{dn_{o}}{dt} = - C {n_o}^2,
\end{eqnarray}
where $n_o$ is the orthoexciton density and the constant $C$ is evaluated to be of the order of 
$\sim 5 \times 10^{-16} \ \mathrm{cm^{3}/ns}.$
From the spectral area of its Lyman absorption line, 
we estimate the orthoexciton density
\begin{eqnarray}
n _{o} = \frac{\hbar c \sqrt{\varepsilon _b}}{4 \pi ^2\Delta E_{1s \text{-}2p}|\mu _{1s \text{-}2p}|^2}
\int_0^{\infty}\Delta \alpha _o(E)\text{d}E .
\end{eqnarray}
Where $\Delta E_{1s \text{-}2p}$ and $\mu _{1s \text{-}2p}$ are 
the transition energy and dipole moment of $1s$-$2p$ transition. 
The background dielectric function $\varepsilon _b$ in the frequency of Lyman transition
is estimated to be about 7. In our previous paper \cite{kubouchi}, 
we took a $1s$-$2p$ dipole moment of 6.3 $e$\AA$\ $by direct analogy with hydrogen atom \cite{artoni}. 
For the yellow series of excitons in Cu$_2$O, we need to take into account 
the central cell correction effects which reduces the overlap 
between $1s$ and $2p$ exciton wave function yielding the revised dipole moment of 1.64 $e$\AA \cite{baym,jorger}.  
The exponential values obtained from the experiment at different pump densities are plotted in Fig. \ref{figure5}(b) 
as a function of the orthoexciton density estimated from the above formula.
From the slope shown in Fig. \ref{figure5}(b), we obtain the constant 
$C_{\text{exp}} = (2.8 \pm 0.4) \times 10^{-15} \ \text{cm}^3$/ns, a factor 6 larger 
than the prediction of Ref.\cite{prb61_16619}. The spin exchange mechanism therefore provides 
a convincing scenario to explain the rapid orthoexciton-paraexciton conversion occurring 
on a 100 ps time scale, when the orthoexciton density exceeds 10$^{15}$ cm$^{-3}$. 

We finally address the initial kinetics, when a narrow line at 129 meV is seen at an orthoexciton density
$< 10^{15}$ cm$^{-3}$ [Fig. \ref{figure4}(b)].
The spin exchange process is not expected to contribute significantly in such a low density regime. 
We identify the fast response as being due to the contribution of the $1s$-$3p$ orthoexciton transition,
which accidentally lies close to the $1s$-$2p$ paraexciton line.
In order to confirm this interpretation, excperiments should be performed with better spectral resolution at early times, 
in order to resolve the higher order term $1s$-$4p$ of the Lyman series.

In conclusion, we have demonstrated a scheme to detect the build up of paraexcitons following the creation of 
an ultracold orthoexciton population in Cu$_2$O. 
Orthoexciton-paraexciton conversion by spin exchange between pairs of orthoexcitons has been detected. 
Metastable biexcitons could resonantly enhance the scattering similar to the Feshbach resonance of cold atoms.
The observed enhanced production of spin forbidden paraexcitons from cold orthoexcitons provides 
a unique opportunity to reach BEC states of excitons.  

The authors are grateful to S. Nobuki for the experimental support. 
The authors are also grateful to T. Tayagaki, N. Naka, and Yu. P. Svirko for stimulating discussions. 
This work is partly supported by the Grant-in-Aid for Scientific Research (S) from 
Japan Society for the Promotion of Science (JSPS).

\end{document}